\documentclass{article}
\usepackage[T1]{fontenc} % add special characters (e.g., umlaute)
\usepackage[utf8]{inputenc} % set utf-8 as default input encoding
\usepackage{ismir,amsmath,cite,url}
\usepackage{graphicx}
\usepackage{color}
\usepackage{comment}
\usepackage{cleveref}
\usepackage{booktabs}
\def\lbd{}	% Flag to use correct LBD settings in the paper, please do not modify this line

\usepackage{lineno}
\newcommand{\eg}{e.\,g.\,}
\newcommand{\ie}{i.\,e.\,}

\title{Can MusicGen create training data for MIR tasks?}
\multauthor
{Nadine Kroher$^1$ \hspace{1cm} Helena Cuesta$^2$ \hspace{1cm} Aggelos Pikrakis$^{1,3}$\vspace{-0.3cm}} { \\
  $^1$ Time Machine Capital 2 Ltd., London, UK\\
$^2$ DAACI Ltd., London, UK\\
$^3$ Dept. of Informatics, University of Piraeus, Greece\\
{\tt\small \{nadine,\,aggelos\}@tmc2.ai, helena@daaci.com}
}

\def\authorname{N. Kroher, H. Cuesta, and A. Pikrakis}

\begin{document}

\maketitle
\begin{abstract}
We are investigating the broader concept of using AI-based generative music systems to generate training data for Music Information Retrieval (MIR) tasks. To kick off this line of work, we ran an initial experiment in which we trained a genre classifier on a fully artificial music dataset created with MusicGen. We constructed over 50\,000 genre-conditioned textual descriptions and generated a collection of music excerpts that covers five musical genres. Our preliminary results show that the proposed model can learn genre-specific characteristics from artificial music tracks that generalise well to real-world music recordings. 

\end{abstract}
\section{Introduction}
\label{sec:intro}
Machine learning systems for audio-based Music Information Retrieval (MIR) tasks, in particular those that rely on Deep Learning (DL), require large amounts of annotated music data, \eg, audio tracks annotated with chords, genre, or instrument labels, for training. Manual labelling of such large quantities at a high quality is time-consuming and expensive and large-scale crowd-sourced annotation efforts often result in noisy data. In addition, recent discussions around the legal aspects of training generative models on copyrighted data has also expanded to analysis models and, in the case of music, there still exists uncertainty around the legality of training classifiers on music without the explicit permission from the rights holders (which is virtually impossible to obtain for smaller players in the field).
We see an opportunity in making use of the recent advancements in generative music systems such as MusicGen~\cite{Copet23_MusicGen_arxiv}, Riffusion~\cite{ForsgrenMartiros22_Riffusion}, JEN-1~\cite{Li23_JEN1-arxiv}, or MusicLM~\cite{Agostinelli23_musiclm_arxiv}, by employing them to create artificial data which can then be used to train MIR systems, such as music tagging engines. More specifically, we select MusicGen, which is available as an open source module and has been trained on data which is covered by a legal agreement.  

\noindent In this work, we describe a first experiment which is part of a larger initiative to investigate how generative AI systems can be leveraged to build novel artificial datasets for MIR tasks. 
This initial study focuses on genre tagging, which aims at assigning a genre tag to a recording based on its musical characteristics~\cite{tzanetakis2002musical}. In the scope of this experiment, we consider five musical genres: ``Electronica'', ``Funk'', ``Orchestral'', ``Pop'', and ``Rock''.

\noindent Below, we first describe the process of creating text prompts (\Cref{sec:data}). We then outline the architecture of the genre classifier (\Cref{sec:dl}) and share the results we obtain (\Cref{sec:eval}) on a small benchmark evaluation set. Finally, we finish with a few observations and suggestions for future work (\Cref{sec:further}).

%\noindent Training DL models on artificially generated data is not novel---it has been done before even in the MIR field. For instance, synthesising music given a score representation of a piece, or a MIDI file. However, using generative AI solutions for music goes beyond because it involves not only generating artificial \emph{sound} but creating the actual piece of music.

%
\section{Training Data Generation}
\label{sec:data}
\begin{figure}
\resizebox{0.85\columnwidth}{!}{%
 \centerline{
 \includegraphics[width=0.9\columnwidth]{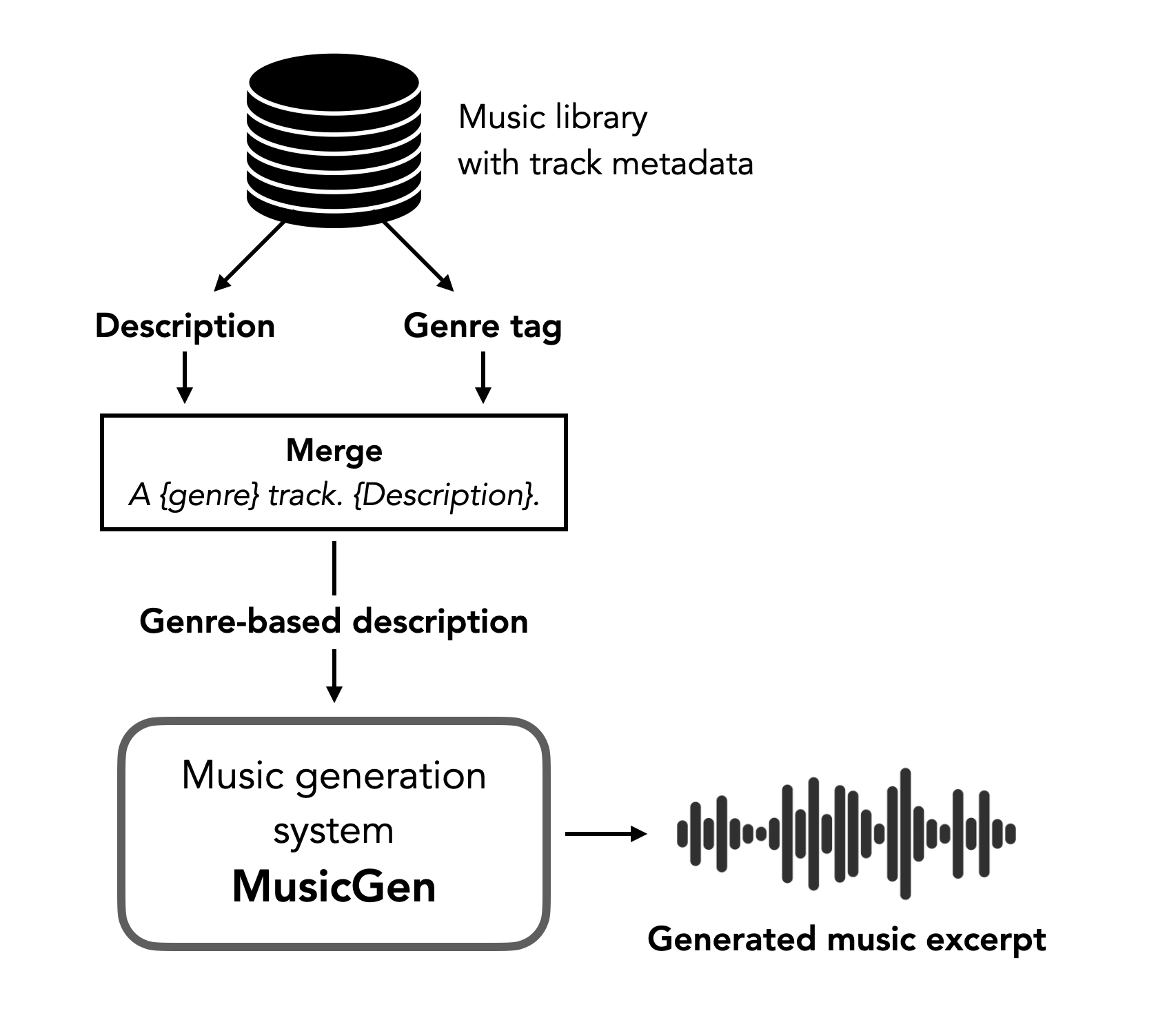}}
 }
 \caption{Diagram of the artificial data generation process.}

 \label{fig:data}
\end{figure}
We leverage MusicGen to create a dataset of 10-second music excerpts. The process is illustrated in \Cref{fig:data}.
We use the small version of the generative model, \ie, \texttt{musicgen-small}, via the AudioCraft library\footnote{\url{https://audiocraft.metademolab.com}} and feed it with text prompts that describe the desired tracks.
More specifically, we compile a set of track descriptions from a music library partner and modify them to specify the genre explicitly. To this end, we parse the track metadata to obtain its associated genre tag and textual description, and combine them into a prompt for MusicGen as follows:

\begin{center}
\vspace{-0.45cm}``A \{\emph{genre-tag}\} track. \{\emph{track-description}\}.''\vspace{-0.2cm}
\end{center}

\noindent where \emph{genre-tag} is one of \{``Electronica'', ``Funk'', ``Orchestral'', ``Pop'', ``Rock''\} and \emph{track-description} is the textual description as given in the track metadata. For example, a pop track prompt from our collection is:
\emph{``A Pop track. Reflective guitars with driving electro drums and bass.''} 

\noindent Following this process, we build an initial collection of text prompts for the five genre tags. Then, we correct the class imbalance by generating multiple tracks from the same prompt, having previously ensured that multiple generations based on the same prompt yield significantly different results. With the final list of track descriptions, we run MusicGen to create a dataset of 57\,562 music excerpts evenly distributed among the genre classes.
The resulting audio files are monophonic and sampled at 32\,000 Hz.
We split our dataset into training (90\%) and validation (10\%) sets for model training.

\section{Genre tagging model}
\label{sec:dl}
Our pipeline consists of a pre-trained audio embedding model which feeds into a shallow neural network to predict a single musical genre tag given an input audio recording.
More specifically, we run a forward pass through the PaSST model~\cite{Koutini22_PASST_Interspeech}, which was trained on AudioSet~\cite{Gemmeke17_AudioSet_ICASSP}, and use the output of the layer before the classifier head as an audio representation. This framework has previously shown promising results on a variety of MIR tasks, including music auto-tagging~\cite{DingL23_KDPasst_ISMIR}.

\noindent During training, we first compute the 768-dimensional PaSST embedding for each audio file in the artificial music dataset. 
Then, the embeddings are fed to a classifier with one single dense layer of 128 units and ReLU activation. 
The output is subsequently passed to the final layer with five nodes and softmax activation, where each output node corresponds to one of the genre classes defined above.
The model is trained using the Adam optimiser, an initial learning rate of 1e-4, and class weights to compensate for a minor remaining class imbalance. We implement an early-stopping mechanism based on the validation loss with a patience of 5 epochs. 
\section{Evaluation and results}
\label{sec:eval}
\begin{table}[t]
    \centering
    \resizebox{0.75\columnwidth}{!}{%
    \begin{tabular}{cccc}
    \toprule
       \textbf{Genre} & \textbf{Precision} & \textbf{Recall} & \textbf{F1-Score} \\ \toprule
       Electronica & 89\% & 85\% & 87\% \\ \midrule
       Funk & 90\% & 90\% & 90\% \\ \midrule
       Orchestral & 100\% & 100\% & 100\% \\ \midrule
       Pop & 85\% & 85\% & 85\% \\ \midrule
       Rock & 95\% & 100\% & 98\% \\ \bottomrule
       \toprule
       \textbf{Accuracy} & 92\% & & \\ \bottomrule
       
    \end{tabular}
}
    \caption{Classification results on the benchmark dataset.}
    \label{tab:rwresults}

\end{table}
The model reaches the best validation loss after 5 epochs and yields a classification accuracy of 84.6\% on the validation split of the artificial dataset. 

\noindent In order to assess the generalisation capabilities of the model to real-world data, we evaluate its performance on a small manually annotated benchmark dataset containing 20 commercial music recordings per genre. All tracks are taken from an in-house music library and were manually labelled by music experts. For classification, we again first extract the PaSST embeddings and then run a forward pass through the trained classifier. 
\Cref{tab:rwresults} shows precision, recall, and F1-score per class as well as the overall accuracy obtained on this real-world music dataset. Although the study covers only a limited range of musical genres, these results suggest that the model successfully learned features and patterns from artificial data that generalise well to those found in real music recordings. These encouraging findings motivate us to further explore the use of AI-based generative music systems for the creation of data for MIR tasks, always considering some of the limitations that we found and discuss in the next section.

\section{Observations and further work}
\label{sec:further}
This initial proof-of-concept study is limited to a small taxonomy of five musical genres for which we verified that the generative model produces convincing results. In order to extend this method to a broader set of classes, some current limitations of MusicGen need to be overcome. 
In particular, the generative model, at least the version used in this study, does not appear to generate vocals, even if specifically stated in the prompt. This is particularly challenging for genres that are heavily focused on vocals, \ie, ``hip hop''. 
In addition, the quality of the generated samples appears to vary across styles. We observed that prompts referring to world music styles (\ie, ``Fado'') or rather uncommon genres (\ie, ``Krautrock'') often yield irrelevant output. We speculate that this could happen because certain genres may have been underrepresented in the training data. Since MusicGen was trained on a private collection with an unknown distribution across musical genres or instrumentation, we cannot ensure that the generated tracks are not biased towards a few specific genres, instruments, or music cultures.
Following this initial experiment, we will continue to explore the concept of using generative music systems in the context of training data creation. More specifically, we will explore more advanced prompt engineering strategies and, rather than training on artificial data only, explore domain adaptation methods \cite{farahani2021brief} to ensure generalisation capabilities by training on large amounts of artificial data and small amounts of real-world data simultaneously.

\section{Conclusion}
\label{sec:conclusion}
As a first step towards leveraging AI-based generative music systems for training data generation, we created a large collection of artificial music clips using MusicGen and trained a genre classifier that shows generalisation capabilities to real-world data. We believe that, with some additional research, this approach can scale to larger taxonomies and other MIR tasks.

% For bibtex users:
\bibliography{ISMIR2023_lbd}

\begin{thebibliography}{1}
\providecommand{\url}[1]{#1}
\csname url@samestyle\endcsname
\providecommand{\newblock}{\relax}
\providecommand{\bibinfo}[2]{#2}
\providecommand{\BIBentrySTDinterwordspacing}{\spaceskip=0pt\relax}
\providecommand{\BIBentryALTinterwordstretchfactor}{4}
\providecommand{\BIBentryALTinterwordspacing}{\spaceskip=\fontdimen2\font plus
\BIBentryALTinterwordstretchfactor\fontdimen3\font minus \fontdimen4\font\relax}
\providecommand{\BIBforeignlanguage}[2]{{%
\expandafter\ifx\csname l@#1\endcsname\relax
\typeout{** WARNING: IEEEtran.bst: No hyphenation pattern has been}%
\typeout{** loaded for the language `#1'. Using the pattern for}%
\typeout{** the default language instead.}%
\else
\language=\csname l@#1\endcsname
\fi
#2}}
\providecommand{\BIBdecl}{\relax}
\BIBdecl

\bibitem{Copet23_MusicGen_arxiv}
J.~Copet, F.~Kreuk \emph{et~al.}, ``{Simple and Controllable Music Generation},'' \emph{arXiv preprint arXiv:2306.05284}, 2023.

\bibitem{ForsgrenMartiros22_Riffusion}
\BIBentryALTinterwordspacing
S.~Forsgren and H.~Martiros, ``{Riffusion - Stable diffusion for real-time music generation},'' 2022. [Online]. Available: \url{https://riffusion.com/about}
\BIBentrySTDinterwordspacing

\bibitem{Li23_JEN1-arxiv}
P.~Li \emph{et~al.}, ``{JEN-1: Text-Guided Universal Music Generation with Omnidirectional Diffusion Models},'' \emph{arXiv preprint arXiV:2308.04729}, 2023.

\bibitem{Agostinelli23_musiclm_arxiv}
A.~Agostinelli, T.~I. Denk \emph{et~al.}, ``Music{LM: Generating Music From Text},'' \emph{arXiv preprint arXiv:2301.11325}, 2023.

\bibitem{tzanetakis2002musical}
G.~Tzanetakis and P.~Cook, ``{Musical Genre Classification of Audio Signals},'' \emph{IEEE Transactions on speech and audio processing}, vol.~10, no.~5, pp. 293--302, 2002.

\bibitem{Koutini22_PASST_Interspeech}
K.~Koutini \emph{et~al.}, ``{Efficient Training of Audio Transformers with Patchout},'' in \emph{Proc. of Interspeech}, 2022, pp. 2753--2757.

\bibitem{Gemmeke17_AudioSet_ICASSP}
J.~F. Gemmeke \emph{et~al.}, ``{Audio Set: An ontology and human-labeled dataset for audio events},'' in \emph{IEEE International Conference on Acoustics, Speech and Signal Processing (ICASSP)}.\hskip 1em plus 0.5em minus 0.4em\relax IEEE, 2017, pp. 776--780.

\bibitem{DingL23_KDPasst_ISMIR}
Y.~Ding and A.~Lerch, ``{Audio Embeddings as Teachers for Music Classification},'' in \emph{Proc. of the International Society for Music Information Retrieval Conference (ISMIR).}, 2023.

\bibitem{farahani2021brief}
A.~Farahani, S.~Voghoei, K.~Rasheed, and H.~R. Arabnia, ``{A Brief Review of Domain Adaptation},'' in \emph{{Advances in Data Science and Information Engineering}}.\hskip 1em plus 0.5em minus 0.4em\relax Springer International Publishing, 2021, pp. 877--894.

\end{thebibliography}

\end{document}